\documentclass[12pt]{iopart}
\usepackage{iopams} 

\def\p{{\boldsymbol p}}
\def\q{{\boldsymbol q}}

\def\k{{\boldsymbol k}}

\def\x{{\boldsymbol x}}

\begin{document}

\title{Initial state effects in the Color Glass Condensate}

\author{Francois Gelis\dag\ and Raju Venugopalan\ddag  
}

\address{\dag\ Service de Physique Th\'eorique,
  CEA/Saclay,
  91191, Gif-sur-Yvette Cedex, France }

\address{\ddag\ Physics Department,
  Brookhaven National Laboratory,
  Upton, NY 11973, USA }

\begin{abstract}
  The Color Glass Condensate provides a systematic power counting of
  initial  state effects in high energy QCD.  We briefly
  discuss in this framework quark and gluon production in high energy
  collisions.
\end{abstract}

\section{Introduction}
The Color Glass Condensate (CGC) is an effective theory of the
hadronic wave-function at high energies. For a review and references, see Ref.~\cite{IancuV}. Key features,
as summarized in the McLerran-Venugopalan model (MV-model)~\cite{MV},
are as follows. The dynamical degrees of freedom are small $x$ partons
with occupation numbers of {\cal O}($\alpha_s^{-1}$) best described by
the classical gauge fields $A^\mu_a$. The large $x$ partons are static
light-cone sources of color charge represented by the two dimensional
color charge density $\rho_a(\x_\perp)$. The classical field of the
nucleus obeys the Yang-Mills equations $[D_\mu,F^{\mu\nu}]_a =
\delta^{\nu+}\delta(x^-)\,\rho_a(\x_\perp)$.  The sources $\rho_a$ are
described by the gauge invariant weight function $W_{x_0}[\rho_a]$
where $x_0$ separates ``fields" and ``sources". As $x_0$ decreases,
fields are transformed into sources -- the weight functional obeys a
renormalization group equation (the JIMWLK equation)
describing its evolution with $x_0$. The evolution of the system is
characterized by the saturation scale $Q_s(x) \gg\Lambda_{\rm QCD}$.
One determines {\it a posteriori} that $\alpha_s(Q_s)\ll1$.
Observables $O[A^\mu]$ are computed in the classical field for fixed
$\rho_a$ and then averaged over $W_{x_0}[\rho_a]$ to obtain the gauge
invariant expectation value: $\big<O\big> = \int [D\rho_a]\,
O[A^\mu(\rho_a)]\, W_{x_0}[\rho_a]$.

In the CGC framework, the ratio of the density $\rho$ of hard parton
color charges to the transverse momentum scale $k_\perp$ of interest
($\rho/k_\perp^2$) enables a systematic power counting of initial and
final state effects in the scattering of both dilute and dense
partonic systems~\cite{IancuV}. These will now be discussed.

\section{Gluon and Quark production in the dilute/pp regime ($\rho_{p1}/k_\perp^2\,,\rho_{p2}/k_\perp^2 \ll 1$)}
Inclusive gluon production in the CGC is computed by solving the
Yang-Mills equations $[D_\mu,F^{\mu\nu}]_a = J^{\nu}_a$, where $J^\nu
= \delta(x^-)\delta^{\nu +}\rho_{p1} + \delta(x^+)\delta^{\nu
  -}\rho_{p2}$, with initial conditions given by the Yang-Mills fields
of the two nuclei before the collision. To lowest order in
$\rho_{p1}/k_\perp^2$ and $\rho_{p2}/k_\perp^2$, one can compute
inclusive gluon production analytically. This was first done in the
$A^\tau=0$ gauge ~\cite{KMW} and subsequently in the Lorenz gauge
$\partial_\mu A^\mu=0$~\cite{KovchegovR}.  The result at this order is
$k_\perp$-factorizable into the product of the unintegrated gluon
distributions in the two projectiles~\footnote{This quantity is not the usual unintegrated distribution but is closely related. See Ref.~\cite{BlaizotGV2} for a discussion.}. The comparison of this result to the
collinear pQCD $gg\rightarrow gg$ process and the $k_\perp$-factorized
$gg\rightarrow g$ was performed in ref.~\cite{GyulassyM}. This result
for gluon production is substantially modified, as we shall discuss
later, by high parton density effects in the projectiles.

The result for inclusive quark pair production can be expressed in
$k_\perp$-factorized form as~\cite{GelisV}
\begin{eqnarray}
\frac{d\sigma_1}{dy_p dy_q d^2\p_\perp d^2\q_\perp}
&=&\frac{1}{(2\pi)^6 d_{_{A}}^2} 
\int\frac{d^2 \k_{1\perp}}{(2\pi)^2}\frac{d^2 \k_{2\perp}}{(2\pi)^2}
\delta(\k_{1\perp}+\k_{2\perp}-\p_\perp-\q_\perp)\,\nonumber \\
& & \qquad\times\varphi_1(k_{1\perp}) \varphi_2(k_{2\perp})
\frac{{\rm Tr}\,\left(\left|m^{-+}_{ab}(k_1,k_2;\q,\p)\right|^2\right)}
{k_{1\perp}^2 k_{2\perp}^2} \; ,
\label{eq:ktfac1}
\end{eqnarray}
where $\varphi_1$ and $\varphi_2$ are the unintegrated gluon
distributions in the projectile and target respectively (with the
gluon distribution defined as $xG(x,Q^2) = \int_0^{Q^2} d(k_\perp^2)\,
\varphi(x,k_\perp)$).  The matrix element ${\rm Tr}\,\big
|m^{-+}_{ab}(k_1,k_2;q,p)\big|^2$ is identical to the result derived
in the $k_\perp$-fac\-to\-ri\-za\-tion approach, which
has been ap\-plied ex\-ten\-si\-ve\-ly to stu\-dy hea\-vy quark
pro\-duc\-tion at collider energies~\cite{Shabelski}. In the limit
$k_{1\perp}\,,k_{2\perp}\rightarrow 0$, ${{\rm
    Tr}\,\big|m^{-+}_{ab}(k_1,k_2;q,p)\big|^2} /{k_{1\perp}^2
  k_{2\perp}^2}$ is well defined -- after integration over the
azimuthal angles in eq.~\ref{eq:ktfac1}, one obtains the usual matrix
element $|{\cal M}|_{gg\rightarrow q\bar q}^2$, recovering the lowest
order pQCD collinear factorization result.

\section{Gluon and Quark production in the semi-dense/pA regime ($\rho_{p}/k_\perp^2\ll 1\,,\rho_{_A}/k_\perp^2\sim 1$)}
Here one solves the Yang-Mills equations $[D_\mu,F^{\mu\nu}]=J^\nu$
(with $J^{\nu}=\delta^{\nu+}\delta(x^-)\,\rho_p(\x_\perp)+\delta^{\nu
  -}\delta(x^+)\,\rho_{_A}(\x_\perp)$)) to determine the gauge field
produced at lowest order in the proton source density and to all
orders in the nuclear source density. The computations are performed
in Lorenz/covariant gauge $\partial_\mu A^\mu=0$. Gluon production, in
this framework, was first computed by Kovchegov and
Mueller~\cite{KovchegovM}. In ref.~\cite{BlaizotGV1}, the gluon field
produced in pA collisions was computed explicitly. One obtains
\begin{eqnarray}
{ A}^\mu(q)&=&{ A}_{p}^\mu(q)
+\frac{ig}{q^2+iq^+\epsilon}
\int\frac{d^2\k_{1\perp}}{(2\pi)^2}
\Big\{
C_{_{U}}^\mu(\q,\k_{1\perp})\, 
\big[U(\k_{2\perp})-(2\pi)^2\delta(\k_{2\perp})\big]
\nonumber\\
&&\qquad\qquad
+
C_{_{V}}^\mu(\q)\, 
\big[V(\k_{2\perp})-(2\pi)^2\delta(\k_{2\perp})\big]
\Big\}
\frac{{ \rho}_p(\k_{1\perp})}{k_{1\perp}^2}
\; ,
\label{eq:A1infty-final}
\end{eqnarray}
with $\k_2\equiv\q-\k_1$ and $U$ \& $V$ Wilson lines containing all
orders in the nuclear source density $\rho_{_A}$.  The coefficient
functions $C_{_U}^\mu$ and $C_{_V}^\mu$ are simply related to the well
known Lipatov effective vertex $C_{_{L}}^\mu$ through the relation
$C_{_{L}}^\mu=C_{_{U}}^\mu+\frac{1}{2}C_{_{V}}^\mu$.

The path-ordered exponential $U$ is a color matrix arising from the
rotation of the color charge density of the proton source due to
multiple scattering off the nucleus.  The path-ordered exponential $V$
(differing from $U$ by a factor $1/2$ in the argument of the
exponential) arises from the propagation of the produced gluon through
the nucleus. Interestingly, the $V$'s do not appear in the final
result for gluon production. This is because for gluons produced
on-shell one finds remarkably that $C_{_U}\cdot C_{_V}=C_{_V}^2=0$ and
$C_{_U}^2=C_{_L}^2 = 4 k_{1\perp}^2 k_{2\perp}^2/q_\perp^2$. Thus only
bi-linears of the Wilson line $U$ survive in the squared amplitude
that gives the gluon production cross-section. The result is
$k_\perp$-factorizable, except that now one replaces $\varphi_2$ with the
unintegrated nuclear gluon distribution $\varphi_{_A}\propto
\big<U^\dagger U\big>$.  This distribution contains powers
of the usual unintegrated gluon distribution to all orders -- one
recovers the usual unintegrated gluon distribution ($\varphi_2$ in
eq.~\ref{eq:ktfac1}) at large transverse momentum.

Our result in the Lorenz gauge is exactly equivalent to that of
Dumitru \& McLerran in the $A^\tau=0$ gauge~\cite{DumitruM}. The
Cronin effect in proton-nucleus collisions has been studied by us in Ref.~\cite{BlaizotGV1} and
several other authors previously and since and will not be discussed further here.

Quark production can now be computed with the gauge field in
eq.~\ref{eq:A1infty-final}~\cite{BlaizotGV2}. The field is decomposed
into the sum of `regular' terms and `singular' terms; the latter
containing a factor $\delta(x^+)$. The regular terms are the terms
where a) a gluon from the proton interacts with the nucleus and
produces a $q\bar q$-pair after the collision, b) the gluon produces
the pair which then scatters off the nucleus. Naively, these would
appear to be the only possibilities in the high energy limit where the
nucleus is strongly Lorentz contracted. However, in the Lorenz gauge,
one has terms in the gauge field (the `singular' terms, proportional
to $\delta(x^+)$) which correspond to the case where the quark pair is
both produced and re-scatters inside the nucleus! Indeed, the contribution
of this term to the amplitude cancels the contribution of the term
proportional to the $V$'s in the regular part.

Our result for quark pair production, unlike gluon production, is not
$k_\perp$-factorizable. It can however still be written in
$k_\perp$-factorized form as a product of the unintegrated gluon
distribution in the proton times a sum of terms with three
unintegrated distributions, $\varphi^{g,g}_{_{A}}$, $\varphi^{q\bar q,
  g}_{_{A}}$ and $\varphi^{q\bar q,q\bar q}_{_{A}}$. These are
respectively proportional to 2-point, 3-point and 4-point correlators
of the Wilson lines we discussed previously. For instance, the
distribution $\varphi^{q\bar q, g}_{_{A}}$ can be interpreted as the
probability of having a $q\bar q$ pair in the amplitude and a gluon in
the complex conjugate amplitude. For large transverse momenta or large
mass pairs, the 3-point and 4-point distributions collapse to the
unintegrated gluon distribution, and we recover the result for pair
production (eq.~\ref{eq:ktfac1}) in the dilute/pp limit.

Single quark distributions are straightforwardly obtained. Here, the
4-point correlator in $\varphi^{q\bar q,q\bar q}_{_{A}}$ collapses to
a 2-point correlator $\varphi^{q,q}_{_{A}}$ corresponding to a quark
(or anti-quark) in the amplitude and complex conjugate amplitude.

For Gaussian sources, as in the MV-model, these 2,3 and 4-point
functions can be computed exactly as discussed in
ref.~\cite{BlaizotGV2}. Single quark distributions in the MV-model
were recently computed by Tuchin~\cite{Tuchin}.

\section{Gluon and Quark production in the dense/AA regime ($\rho_{_{A1}}/k_\perp^2\, ,\rho_{_{A2}}/k_\perp^2\sim 1$)}
This case is the relevant one for particle production in heavy ion
collisions. It involves solving the Yang-Mills equations to all orders
in the sources of both nuclei. This problem has not been solved
analytically thus far -- $k_\perp$-factorization breaks down completely
here, even for gluon production.

The problem has however been solved numerically~\cite{KNV}.
Non-perturbative formulae are derived relating (for collisions of
identical nuclei) the saturation scale $Q_s$ in the nuclear
wave-function to the energy and number distributions of gluons produced
immediately after the collision (on a time scale $\sim 1/Q_s$). The
gluon distribution is infra-red finite and is fit by a massive
Bose-Einstein distribution with a ``temperature'' $T\sim 0.47\,Q_s$
and $m\sim 0.04 \,Q_s$ for $k_\perp \leq 1.5\, Q_s$. For $k_\perp >
1.5\,Q_s$, it is fit by the tree level perturbative form
${Q_s^4\over k_\perp^4}\,\ln(4\pi k_\perp/Q_s)$.

The classical field description is valid as long as the occupation
number $f$ is greater than unity. As the system evolves, it becomes
dilute and the classical description breaks down. Baier, Mueller,
Schiff and Son~\cite{BMSS} estimated this time to be ${\cal O}(
  \alpha_s^{-3/2}Q_s^{-1})$ In their ``bottom-up" scenario,
they suggest that inelastic $2\rightarrow 3$ processes, though
parametrically suppressed, are actually more efficient in driving the
system from the classical stage towards thermalization. The analysis
is valid for very small couplings and suggests that thermalization may
take several fermis to achieve. In this light, the early thermalization
required in some RHIC phenomenology appears puzzling. Arnold, Lenaghan
and Moore have suggested that collective instabilities might
drive the system faster towards equilibrium~\cite{ALM}.

An interesting possibility is suggested by the following. As the
classical field expands, one identifies a scale $\Lambda(\tau_0)$ at
an early time $\tau_0$ which separates high momentum particles from
low momentum classical fields. The high momentum particles scatter off
the fields while the classical fields interact with each other and
with the particles. With time, at an appropriate $\tau_1$, one can
define a new (``coarse graining") scale $\Lambda(\tau_1)$ at which one
re-defines field and particle modes. We have developed an algorithm
for a scalar field theory which implements this dynamical coarse
graining while ensuring energy-momentum conservation~\cite{Weinstock}.
While promising, much work remains to extend this formalism to gauge
theories.


\vspace{0.5cm}
\begin{thebibliography}{9}

\bibitem{IancuV}E.~Iancu and R.~Venugopalan,
%``The color glass condensate and high energy scattering in QCD,''
arXiv:hep-ph/0303204.

\bibitem{MV}L.~D.~McLerran and R.~Venugopalan,
%``Computing quark and gluon distribution functions for very large nuclei,''
Phys.\ Rev.\ D {\bf 49}, 2233 (1994); {\it ibid.}, 3352, (1994); {\it ibid.}, {\bf 50}, 2225 (1994); {\it ibid.}, {\bf 59}:094002, (1999).

\bibitem{KMW}A.~Kovner, L.~D.~McLerran and H.~Weigert,
 %``Gluon production at high transverse momentum in the McLerran-Venugopalan
%model of nuclear structure functions,''
Phys.\ Rev.\ D {\bf 52}, 3809 (1995).

\bibitem{KovchegovR}Y.~V.~Kovchegov and D.~H.~Rischke,
 %``Classical gluon radiation in ultrarelativistic nucleus nucleus
%collisions,''
Phys.\ Rev.\ C {\bf 56}, 1084 (1997).

\bibitem{GyulassyM}M.~Gyulassy and L.~D.~McLerran,
%``Yang-Mills radiation in ultrarelativistic nuclear collisions,''
Phys.\ Rev.\ C {\bf 56}, 2219 (1997)

\bibitem{GelisV}F.~Gelis and R.~Venugopalan,
%``Large mass q anti-q production from the color glass condensate,''
Phys.\ Rev.\ D {\bf 69}, 014019 (2004).

\bibitem{Shabelski}For a review, see M.~G.~Ryskin, Y.~M.~Shabelski and A.~G.~Shuvaev,
%``Heavy quark production in hadron collisions,''
arXiv:hep-ph/0011111.

\bibitem{KovchegovM}Y.~V.~Kovchegov and A.~H.~Mueller,
 %``Gluon production in current nucleus and nucleon nucleus collisions in  a
%quasi-classical approximation,''
Nucl.\ Phys.\ B {\bf 529}, 451 (1998).

\bibitem{BlaizotGV1}J.~P.~Blaizot, F.~Gelis and R.~Venugopalan,
 %``High energy p A collisions in the color glass condensate approach. I: Gluon
%production and the Cronin effect,''
arXiv:hep-ph/0402256.

\bibitem{DumitruM}A.~Dumitru and L.~D.~McLerran,
%``How protons shatter colored glass,''
Nucl.\ Phys.\ A {\bf 700}, 492 (2002).

\bibitem{BlaizotGV2}J.~P.~Blaizot, F.~Gelis and R.~Venugopalan,
 %``High energy p A collisions in the color glass condensate approach. II: Quark
%production,''
arXiv:hep-ph/0402257.

\bibitem{Tuchin}K.~Tuchin,
%``Heavy quark production from color glass condensate at RHIC,''
arXiv:hep-ph/0402298; arXiv:hep-ph/0401022.

\bibitem{KNV}A.~Krasnitz and R.~Venugopalan,
 %``Non-perturbative computation of gluon mini-jet production in nuclear
%collisions at very high energies,''
Nucl.\ Phys.\ B {\bf 557}, 237 (1999); Phys.\ Rev.\ Lett.\  {\bf 84}, 4309 (2000); Phys.\ Rev.\ Lett.\  {\bf 86}, 1717 (2001); A.~Krasnitz, Y.~Nara and R.~Venugopalan,
Phys.\ Rev.\ Lett.\  {\bf 87}, 192302 (2001); T.~Lappi, Phys.\ Rev.\ C {\bf 67}, 054903 (2003).

\bibitem{BMSS}R.~Baier, A.~H.~Mueller, D.~Schiff and D.~T.~Son,
%``'Bottom-up' thermalization in heavy ion collisions,''
Phys.\ Lett.\ B {\bf 502}, 51 (2001); Phys.\ Lett.\ B {\bf 539}, 46 (2002).

\bibitem{ALM}P.~Arnold, J.~Lenaghan and G.~D.~Moore, JHEP {\bf 0308}, 002 (2003).

\bibitem{Weinstock}S. Weinstock et al., in preparation.

\end{thebibliography}
\end{document}